\documentclass[aps,prd,showpacs]{revtex4}

\usepackage{graphicx}
\usepackage{bm}
\makeatother
\begin{document}

\title{The comparison of velocity distribution 
between Adhesion approximation and the Euler-Jeans-Newton model}
\author{Hajime Sotani$^{1}$\thanks{E-mail: sotani@gravity.phys.waseda.ac.jp}, Takayuki Tatekawa$^{1, 2, 3}$\thanks{E-mail: tatekawa@cosmos.phys.ocha.ac.jp}
 }

\affiliation{1. Advanced Research Institute for Science
and Engineering, Waseda University,
3-4-1 Okubo, Shinjuku, Tokyo 169-8555, Japan\\
2. Department of Physics, Ochanomizu University,
2-1-1 Otsuka, Bunkyo, Tokyo 112-8610, Japan\\
3. Department of Physics, Waseda University,
3-4-1 Okubo, Shinjuku, Tokyo 169-8555, Japan}

\date{\today}

%%%%%%%%%%%%%%%%%%%%%%%%%%%%%%%%%%%%%%%%%%%%%%%
\begin{abstract}
For the evolution of density fluctuation in
nonlinear cosmological dynamics, 
adhesion approximation (AA) is proposed as a
phenomenological model, which is especially useful for describing
nonlinear evolution.
However, the origin of the artificial viscosity
in AA is not clarified.
Recently, Buchert and Dom\'{\i}nguez report if
the velocity dispersion of the dust fluid
is regarded as isotropic, it
works on a principle similar to 
viscosity or effective pressure,
and they consider isotropic velocity dispersion
as the origin of the artificial viscosity in AA.
They name their model the Euler-Jeans-Newton (EJN) model.
In this paper, we focus on the velocity distribution in 
AA and the EJN model and examine the time evolution in both models.
We find the behavior of AA differs from that of the
EJN model, i.e., although the peculiar velocity
in the EJN model oscillates,
that in AA is monotonically decelerated due to viscosity without
oscillation. Therefore it is hard to regard
viscosity in AA as effective pressure in the EJN model.
\end{abstract}
%%%%%%%%%%%%%%%%%%%%%%%%%%%%%%%%%%%%%%%%%%%%%%%


\maketitle

%%%%%%%%%%%%%%%%%%%%%%%%%%%%%%%%%%%%%%%%%%%
\section{Introduction}\label{sec:intro}
%%%%%%%%%%%%%%%%%%%%%%%%%%%%%%%%%%%%%%%%%%%

The Lagrangian description for the cosmological fluid
can be usefully applied to
the structure formation scenario. This description provides a
relatively accurate model even in a quasi-linear regime.
Zel'dovich~\cite{zel} proposed
a linear Lagrangian approximation for dust fluid.
This approximation is called the Zel'dovich approximation
(ZA)~\cite{zel,Arnold82,Shandarin89,buchert89,Paddy93,coles,saco,Jones05,Tatekawa05R,Paddy05}.
ZA describes the evolution of density fluctuation better than
the Eulerian approximation
\cite{munshi,sahsha,Yoshisato98,Yoshisato05}.
Although ZA gives an accurate
description until the quasi-linear stage,
ZA cannot describe the model after the formation
of caustics. In ZA, even after the formation of caustics, the fluid elements
keep moving in the direction set up by the initial condition.

In order to proceed with a hydrodynamical description
in which caustics do not form, a qualitative pressure gradient
\cite{Zeldovich82}
and thermal velocity scatter~\cite{Kotok87, Shandarin89}
in a collisionless matter have been discussed.
Similarly, adhesion approximation (AA)~\cite{Gurbatov89} has been proposed,
which is a model based on a nonlinear diffusion equation
(i.e., Burgers's equation~\cite{Burgers40}). In AA, an artificial
viscosity term is added to ZA; thus
we can avoid caustics formation.
The problem of structure formation has been discussed
from the standpoint of AA~\cite{Shandarin89,weinberg90,nusser,Kofman92,msw},
where it is shown that the density divergence does not occur in AA
and that a density distribution close to the N-body simulation can be produced.
However, the origin of the viscosity in AA has not yet been clarified.

Buchert and Dom\'{\i}nguez~\cite{budo} discussed the effect
of velocity dispersion using the collisionless Boltzmann equation~\cite{BT}.
They argued that models of a large-scale structure should
be constructed by a flow describing the average
motion of a multi-stream system.
Then they showed that when the velocity dispersion is regarded
as small and isotropic, it produces effective pressure or
viscosity terms, and they consider that the isotropic velocity dispersion corresponds
to the origin of the artificial viscosity in AA.
Furthermore, in consideration of kinematic
theory, they derived the
relation between mass density $\rho$ and pressure $P$, i.e.,
an equation of state.
Buchert {\it et al.}~\cite{bdp} also showed how the viscosity term
or the effective pressure of a fluid is generated,
assuming that the peculiar acceleration is
parallel to the peculiar velocity.
Recently Buchert and Dom\'{\i}nguez~\cite{Buchert05}
provided an evaluation of the current status of adhesive
gravitational clustering, which includes the above discussion,
and they tried to improve past models. In their paper, they named
their approach the Euler-Jeans-Newton (EJN) model.
On the other hand, Dom\'{\i}nguez~\cite{domi00,domi0106} proposed another approach.
In these papers he clarified that a hydrodynamic
formulation is obtained via a spatial coarse-graining
in a many-body gravitating system, and that the viscosity term in
AA can be derived by the expansion
of coarse-grained equations. 
This model is named the Small-Size Expansion (SSE) model
~\cite{Buchert05}.

So far, with respect to the correspondence of the viscosity term
with the effective pressure,
and with regard to the extension of the Lagrangian description
to various matter,
the EJN model has been considered.
Actually, Adler and Buchert~\cite{adler} have formulated the
Lagrangian perturbation theory for a barotropic fluid.
Morita and Tatekawa~\cite{moritate}, Tatekawa {\it et al.}
\cite{Tatekawa02},
and Tatekawa~\cite{Tatekawa05A, Tatekawa05B}
solved the Lagrangian perturbation equations for a polytropic fluid.
However, it is still a open problem whether AA could realize the
behavior of the EJN model in the proper way.
It is known that the viscosity in AA
decelerates the peculiar velocity in a dense region
and avoids the formation of the caustic,
while the effective pressure in the EJN model
also decelerates the motion of the fluid.
But the fluid in the EJN model not only would decelerate
but also might bounce in a dense region
due to the effective pressure,
which is known as the Jeans instability for cosmological fluid.
In other words, we consider that
AA would not realize the behavior of the EJN model. 

In this paper, to compare AA and the EJN model,
we especially analyze the peculiar
velocity in a cylindrical collapse model.
Although we already compared AA with
the EJN model with respect to density fluctuation
in a past paper~\cite{Tatekawa04}, because we used an
explicit method for solving partial differential equations,
the accuracy of the numerical
calculation in the dense region was not good. Therefore we could not observe
the Jeans instability but, rather, the numerical instability.
In this paper, instead of the explicit method, we apply
the iterated Crank-Nicholson method~\cite{Teukolsky}.
This method resolves the difficulty of solving the resulting implicit
algebraic equations in the original Crank-Nicholson method
and preserves the good stability properties.

As our analyses show, although the peculiar velocity in AA decelerates
due to the viscosity, the velocity does not oscillate. On the other hand,
as a preliminary expectation,
the peculiar velocity in the EJN model decelerates and also oscillates.
Although the tendency of the evolution of the peculiar velocity would
depend on the viscous parameter or the Jeans length, the behavior of AA is
obviously different from that of the EJN model. We notice that while both AA and the EJN model
certainly describe the quasi-nonlinear evolution well, the detail of
the evolution is different.
When we take a large value for the Jeans length, the fluid bounces
to the outside. On the other hand, when we take a small value for
the Jeans length, although the collapse of the cylindrical matter
decelerates, a caustic might finally form at the center.
Therefore it is problematic that the viscous term in AA is explained
as an effect similar to the effective pressure term in the EJN model.
For an explanation of the viscous term in AA, we will have to
analyze more Lagrangian models.

This paper is organized as follows.
In Sec.~\ref{sec:Lagrangian}, we present Lagrangian perturbative solutions
in the Einstein-de Sitter (E-dS) universe.
First, we show linear perturbative solutions
for dust fluid in Sec.~\ref{subsec:dust}, and
in Sec.~\ref{subsec:adhesion} we mention the problem of ZA and show
the solution of AA. Then we explain the EJN model in
Sec.~\ref{subsec:pressure}.
In Sec.~\ref{sec:compari} we compare the evolution of 
the peculiar velocity in AA with that in the EJN model.
Finally in Sec.~\ref{sec:discuss} we discuss our results
and state our conclusions.

%%%%%%%%%%%%%%%%%%%%%%%%%%%%%%%%%%%%%%%%%%%%%%%%%%%%%%

%%%%%%%%%%%%%%%%%%%%%%%%%%%%%%%%%%%%%%%%%%%%%%%%%%%%%%%%%%%%%%%%%%%%%%%%%%%%%%%%%%%%%%
\section{The Lagrangian description for the cosmological fluid}\label{sec:Lagrangian}
%%%%%%%%%%%%%%%%%%%%%%%%%%%%%%%%%%%%%%%%%%%%%%%%%%%%%%%%%%%%%%%%%%%%%%%%%%%%%%%%%%%%%%

In this section, we briefly present perturbative solutions in the Lagrangian
description.
In Newtonian cosmology,
to introduce cosmic expansion, we adopt the coordinate transformation
from physical coordinates $\bm{r}$ to comoving coordinates $\bm{x}$, such as
$\bm{x} = \bm{r}/a(t)$, where $a(t)$ is a scale factor.
In Lagrangian hydrodynamics,
the comoving coordinates $\bm{x}$ of the fluid elements are
represented in terms of Lagrangian coordinates $\bm{q}$ as
\begin{equation} \label{eqn:x=q+s}
\bm{r} = a \bm{x} = a \left ( \bm{q} + \bm{s} (\bm{q},t) 
\right ) \,,
\end{equation}
where $\bm{s}$ denotes the Lagrangian displacement vector
due to the presence of inhomogeneities.
With the Jacobian of the coordinate transformation from
$\bm{x}$ to $\bm{q}$, $J \equiv \det (\partial x_i / \partial q_j)
= \det (\delta_{ij} + \partial s_i / \partial q_j)$,
the mass density is described exactly as
\begin{equation}\label{eqn:exactrho}
\rho = \rho_{\rm b} J^{-1} \,,
\end{equation}
where $\rho_{\rm b}$ means background average density.
Furthermore we can decompose $\bm{s}$ into the longitudinal
and the transverse modes, i.e.,
$\bm{s} = \nabla_{\bm{q}} S
+ \bm{S}^{\rm T}$ with
$\nabla_{\bm{q}} \cdot \bm{S}^{\rm T}=0$.
In this paper, we consider only the longitudinal mode for simplicity.
The evolution equation for the longitudinal mode is written as follows
~\cite{adler,moritate}:
\begin{eqnarray} \label{eqn:P-full}
 \nabla_x \cdot \left ( \ddot{\bm{s}} + 2 \frac{\dot{a}}{a}
 \dot{\bm{s}} - \frac{\kappa \gamma \rho_b^{\gamma-1}}{a^2}
 J^{-\gamma} \nabla_x J \right )
 &=& -4 \pi G \rho_b (J^{-1} -1) \,,
\end{eqnarray}
where the dot above the variables denotes the partial derivative with respect to $t$.
In general, it is very difficult to solve this equation for such
reasons as the coordinate transformation or non-locality.
In order to avoid this difficulty, we apply the perturbative approach and
impose symmetry in Eq.~(\ref{eqn:P-full}).
Particularly in this paper we consider cylindrical symmetry, and
the evolution equation in a cylindrical symmetric model is given
in Appendix~\ref{sec:Sphe-eqs}.

%%%%%%%%%%%%%%%%%%%%%%%%%%%%%%%%%%%%%%%%%%%%%%%%%%%%%%%%%%%%%%%%%%%%%%%%%%%%%
\subsection{The Lagrangian perturbation for dust fluid}\label{subsec:dust}
%%%%%%%%%%%%%%%%%%%%%%%%%%%%%%%%%%%%%%%%%%%%%%%%%%%%%%%%%%%%%%%%%%%%%%%%%%%%%

Zel'dovich derived a first-order solution of the longitudinal mode
for dust fluid~\cite{zel}. In the Friedmann Universe model,
the solutions are formally written as follows:
\begin{equation} \label{eqn:sol-ZA}
S (\bm{q}, t) = D_+(t) S_+ (\bm{q}) + D_-(t) S_- (\bm{q}) \,,
\end{equation}
where $D_+(t)$ and $D_-(t)$ mean the growing factor and
the decaying factor, respectively.
This first-order approximation is called the Zel'dovich
approximation (ZA). In the case of the Einstein-de Sitter Universe,
$D_+(t)$ and $D_-(t)$ are described as
\begin{eqnarray}
D_+(t) & \propto & t^{2/3} \,, \label{eqn:ZA-grow} \\
D_-(t) & \propto & t^{-1} \,.
\end{eqnarray}
Especially when we consider
the plane-symmetric case, ZA gives exact solutions
\cite{Doroshkevich73,Arnold82}.
%%%%%%%%%%%%%%%%%%%%%%%%%%%%%%%%%%%%%%%%%%%%%%%%%%%%%%%%%%%%%%%%
\subsection{Adhesion approximation (AA)}\label{subsec:adhesion}
%%%%%%%%%%%%%%%%%%%%%%%%%%%%%%%%%%%%%%%%%%%%%%%%%%%%%%%%%%%%%%%%

Although the Lagrangian approximation gives an accurate
description until a quasi-linear regime develops,
it cannot describe the model after the formation of caustics.
After that, the nonlinear structure diffuses because
the fluid elements
keep moving in the direction set up by the initial condition.
In order to avoid caustics formation,
the adhesion approximation (AA)~\cite{Gurbatov89} was
proposed, which is a model based on a nonlinear diffusion equation
(Burgers's equation~\cite{Burgers40}).
In AA, an artificial
viscosity term is added to ZA.
In ZA, the equation for ``peculiar velocity'' $\bm{u}$
is written as follows:
\begin{eqnarray}
\frac{\partial \bm{u}}{\partial D_+}
 + (\bm{u} \cdot \nabla_x) \bm{u}
&=& 0 \,, \label{ZA-equation} \\
\bm{u} \equiv \frac{\partial \bm{x}}{\partial D_+}
 = \frac{\dot{\bm{x}}}{\dot{D_+}} \label{eqn:def-u} \,,
\end{eqnarray}
where $D_+$ is the growing factor in ZA.
To avoid caustics formation, in AA
we add an artificial viscosity term to the right side
of the Eq.~(\ref{ZA-equation}), i.e.,
\begin{equation} \label{eqn:adhesion}
\frac{\partial \bm{u}}{\partial D_+} + (\bm{u} \cdot \nabla_x) \bm{u}
= \nu \nabla_x^2 \bm{u} \,.
\end{equation}
Now we introduce the Hopf-Cole transformation~\cite{Hopf50, Cole51} such as
\begin{equation}
\bm{u} = -\nu \nabla_x \left ( \log \theta(\bm{x}, D_+) \right ) \,;
\end{equation}
then the Eq.~(\ref{eqn:adhesion}) is changed to a diffusion
equation:
\begin{equation} \label{eqn:diffusion}
\frac{\partial \theta}{\partial D_+} = \nu \nabla_x^2 \theta \,.
\end{equation}
Meanwhile, by using the inverse Hopf-Cole transformation, we obtain the solution
of Eq.~(\ref{eqn:adhesion}):
\begin{eqnarray}
\bm{x} &=& \bm{q} + \int_{D_0}^{D_+} \bm{u} \left (\bm{x}
 (\bm{q}, D'), D' \right ) {\rm d} D' \,, \label{eqn:sol-AA} \\
\bm{u} &=& \frac{\int {\rm d}^3 \bm{x}' \frac{(\bm{x}-\bm{x}')}{D}
 G(\bm{x}, \bm{x}')}{\int {\rm d}^3 \bm{x}' G(\bm{x}, \bm{x}')} 
 \,, \\
G({\bm x}, {\bm x}') &=& \exp \left [ -\frac{1}{2\nu}
 \left ( \Psi_0 (\bm{x}') + \frac{(\bm{x}-\bm{x}')^2}{2D} 
 \right ) \right ] \,,
\end{eqnarray}
where
\begin{equation}
\nabla_x \Psi_0 (\bm{x}) \equiv \bm{s}_0 \,.
\end{equation}

We consider the case when the viscosity coefficient is quite small
($\nu \rightarrow +0$ ($\nu \ne 0$)). 
Within the limits of a small $\nu$, the analytic solution of
Eq.~(\ref{eqn:adhesion}) is given by
\begin{eqnarray}
\bm{u} (\bm{x}, t) &=& \frac{\sum_{\alpha}
 \left(\frac{\bm{x}-\bm{q}_{\alpha}}
{D_+} \right ) j_{\alpha} \exp \left( -\frac{I_{\alpha}}{2\nu}
  \right ) }
 { \sum_{\alpha} j_{\alpha} \exp \left( -\frac{I_{\alpha}}{2\nu}
 \right ) } \,,
\end{eqnarray}
where $\bm{q}_{\alpha}$ is the Lagrangian points that minimize the
action
\begin{eqnarray}
I_{\alpha} & \equiv & I(\bm{x}, a; \bm{q}_{\alpha})
 = S_0 (\bm{q}_{\alpha}) + \frac{(\bm{x}-\bm{q}_{\alpha})^2}{2a}
 = \mbox{min.} \,, \\
j_{\alpha} & \equiv & \left. \left[ \det \left (\delta_{ij}
 + \frac{\partial^2 S_0}{\partial q_i \partial q_j} \right )
 \right ]^{-1/2} \right |_{\bm{q}=\bm{q}_{\alpha}} \,, \\
S_0 &=& S(\bm{q}, t_0) \,,
\end{eqnarray}
considered as a function of $\bm{q}$ for fixed $\bm{x}$~\cite{Kofman92},
where the Roman character index denotes the Cartesian coordinate.
In this paper, we consider the cylindrical symmetric case. For this case,
we have to change the evolution equation, Eq.~(\ref{eqn:adhesion}), slightly
(see Sec. \ref{sec:compari-AA}).

%%%%%%%%%%%%%%%%%%%%%%%%%%%%%%%%%%%%%%%%%%%%%%%%%%%%
\subsection{The EJN model}\label{subsec:pressure}
%%%%%%%%%%%%%%%%%%%%%%%%%%%%%%%%%%%%%%%%%%%%%%%%%%%%
Although AA seems a good model for avoiding the formation of caustics,
the origin of the modification (or artificial viscosity) is not
clarified. Buchert and Dom\'{\i}nguez~\cite{budo} argued that the effect
of velocity dispersion is important in hindering caustics formation.
They showed that when the velocity dispersion is still
small and can be regarded as isotropic, it behaves as effective
pressure or viscosity terms. Under the consideration
of fluid kinematics, they proposed the effective
equation of state as $p \propto \rho^{5/3}$. Also
Buchert {\it et al.}~\cite{bdp} showed how the viscosity term
is generated by the effective pressure of a fluid
under the assumption that the peculiar acceleration is
parallel to the peculiar velocity.
Moreover Buchert and Dom\'{\i}nguez~\cite{Buchert05} recently
provided an evaluation of the current status of adhesive gravitational
clustering, which is included in the above discussion. In their paper, they named
their approach the Euler-Jeans-Newton (EJN) model.

When we consider the polytropic equation of state $P=\kappa \rho^{\gamma}$,
the first-order solutions for the longitudinal mode can be written in Fourier space
as follows~\cite{moritate};
for $\gamma \ne 4/3$,
\begin{equation}\label{hatSbessel}
\widehat{S}(\bm{K},a) \propto a^{-1/4}
\, \mathcal{J}_{\pm 5/(8-6\gamma)}
\left( \sqrt{\frac{2C_2}{C_1}}
\frac{|\bm{K}|}{|4-3\gamma|}
\, a^{(4-3\gamma)/2} \right) \,,
\end{equation}
where $\widehat{S}$ is the Fourier transformation of $S$,
$\bm{K}$ is Lagrangian wavenumber,
and $\mathcal{J}_{\nu}$ denotes the Bessel function of order $\nu$,
and for $\gamma=4/3$,
\begin{equation}\label{hatS43}
\widehat{S}(\bm{K},a) \propto
a^{-1/4 \pm \sqrt{25/16 - C_2 |\bm{K}|^2 / 2C_1}} \,,
\end{equation}
where $C_1 \equiv 4 \pi G \rho_{\rm b}(a_{\rm in})
\, a_{\rm in}^{\ 3} /3$,
$C_2 \equiv \kappa \gamma \rho_{\rm b}(a_{\rm in})^{\gamma-1}
\, a_{\rm in}^{\ 3(\gamma-1)}$, and $a_{\rm in}$
means the scale factor given as an initial condition.
When we take the limit $\kappa \rightarrow 0$, these solutions
agree with Eq.~(\ref{eqn:sol-ZA}).

In this model, the behavior of the solutions strongly depends on
the relation between the scale of fluctuation and the Jeans scale.
Here we define the Jeans wavenumber as
\begin{equation} \label{eqn:Jeans-WN}
K_{\rm J} \equiv \left(
\frac{4\pi G\rho_{\rm b} a^2}
     {{\rm d} P / {\rm d} \rho (\rho_{\rm b})} \right)^{1/2} \,.
\end{equation}
The Jeans wavenumber, which gives a criterion for
whether a density perturbation with a wavenumber
will grow or decay with oscillation,
depends on time in general. If the polytropic index $\gamma$ is smaller
than $4/3$, all modes become decaying modes and the fluctuation
will disappear. On the other hand, if $\gamma > 4/3$, all density
perturbations will grow to collapse. In the case where $\gamma=4/3$,
the growing and decaying modes coexist at all times.
We can rewrite the first-order solution Eq.~(\ref{hatSbessel}) with
the Jeans wavenumber, i.e.,
\begin{equation}
\widehat{S}(\bm{K},a) \propto a^{-1/4}
\, \mathcal{J}_{\pm 5/(8-6\gamma)}
\left( \frac{\sqrt{6}}{|4-3\gamma|}
\frac{|\bm{K}|}{K_{\rm J}} \right) \,.
\end{equation}

In this paper, we analyze the first-order perturbation.
When we consider cylindrical symmetric models, even if we
deal with only the first-order perturbation,
they cannot be analyzed in Fourier space, and
we need to solve partial differential equations in real space
with a numerical method (also see Appendix \ref{sec:Sphe-eqs}).

%%%%%%%%%%%%%%%%%%%%%%%%%%%%%%%%%%%%%%%%%%%%%%%%%%%%%%%%%%%%%%%%%%%
\section{Time evolution in cylindrical model}\label{sec:compari}
%%%%%%%%%%%%%%%%%%%%%%%%%%%%%%%%%%%%%%%%%%%%%%%%%%%%%%%%%%%%%%%%%%%

\begin{figure}
 \includegraphics{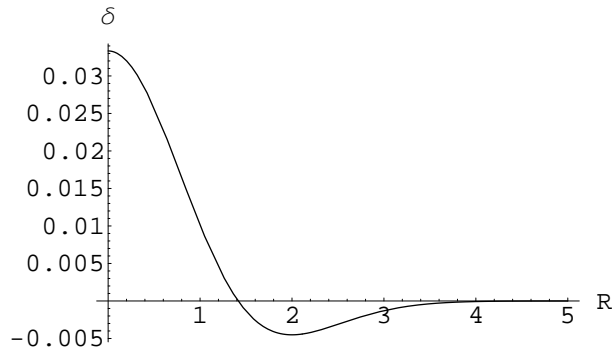}
 \caption{Mexican-hat type model. The average of density fluctuation
 over the whole space becomes zero.}
\label{fig:Mexican}
\end{figure}

For the cylindrical-symmetric case, dust collapse
has been analyzed~\cite{Yoshisato05}.
Here we consider the evolution with ZA, AA, and the EJN models
in the E-dS Universe. Hereafter, we define $R$ and $r$ as Eulerian
and Lagrangian radial coordinates, respectively. At the initial time,
we can identify the Lagrangian coordinate with the Eulerian one,
i.e., $R=r$.
Previous studies considered the collapse and/or evolution
with the top-hat density distribution as the initial condition.
Although the evolution of this model
is easy to compute, the boundary condition
becomes discontinuous.
To avoid a discontinuity of the pressure gradient,
we adopt the Mexican-hat type model (Fig.~\ref{fig:Mexican}):
\begin{equation}
\delta(R) = \varepsilon (2-R^2) e^{-R^2/2} \,,
\end{equation}
where $R$ is Eulerian comoving radial coordinate.
This model has several merits, for example the fluctuation is derived by
the two times differential calculus of Gaussian, i.e.,
\begin{eqnarray}
- \nabla^2 \left ( \varepsilon e^{-R^2/2} \right )
 &=& - \frac{1}{R} \frac{\partial}{\partial R}
\left ( R \frac{\partial}{\partial R}
 \left ( \varepsilon e^{-R^2/2} \right ) \right )
 \nonumber \\
&=& \varepsilon (2-R^2) e^{-R^2/2} \,,
\end{eqnarray}
and the average of density fluctuation over the whole space becomes zero:
\begin{equation}
\int_0^{\infty} 2\pi R \delta(R) {\rm d} R = 0 \,.
\end{equation}
%
%%% Figure %%%
%%%%%%%%%%%%%

For this model, from Eq.~(\ref{eqn:sol-ZA})
the solution of ZA is given as follows:
\begin{equation}
S (a, r) = - a \varepsilon e^{-r^2/2} \,. \label{solution-ZA}
\end{equation}
In our analysis, we set $\varepsilon=1$ due
to an advantage of linear analysis, and
the initial scale factor is set
$a_0=0.0167 (=1/60)$ as the initial condition,
where the initial density fluctuation at $r=0$
becomes $\delta = 1/30$ and
the caustic appears at $a=1$.
On the other hand, for AA and the EJN model
we can determine the initial longitudinal mode $S$ and
the initial ``peculiar velocity'' by preference as the initial conditions.
In this paper they are made equal with those of the growing mode in ZA.
Thus the initial conditions for AA and EJN model are given by
\begin{eqnarray}
S(a_0, r) &=& - a_0 \varepsilon e^{-r^2/2} \,, \\
\left. \frac{\partial S(a, r)}{\partial a} \right |_{a=a_0}
 &=& - \varepsilon e^{-r^2/2} \,.
\end{eqnarray}

%%%%%%%%%%%%%%%%%%%%%%%%%%%%%%%%%%%%%%%%%%%%%%%%%%%%%%%%%%%%%%%%
\subsection{The adhesion approximation}\label{sec:compari-AA}
%%%%%%%%%%%%%%%%%%%%%%%%%%%%%%%%%%%%%%%%%%%%%%%%%%%%%%%%%%%%%%%%

First we consider the evolution in the AA model. 
The evolution of the fluctuation is described by
Eq.~(\ref{eqn:adhesion}). For the cylindrical case,
we slightly change the evolution equation.
When we introduce cylindrical coordinates and
assume the cylindrical symmetry, Burgers's equation
is described as follows~\cite{Nerney}:
\begin{eqnarray}
\partial_{\tau} u + u \partial_{\chi} u &=& \nu \left [\frac{1}{\chi}
 \partial_{\chi} \left(\chi \partial_{\chi} u \right) -
 \frac{u}{\chi^2} \right ] \nonumber \\
 &=& \nu \partial_{\chi} \left (\frac{1}{\chi}
  \partial_{\chi} (\chi u) \right)
 \label{eqn:cylindrical-Burgers} \,,
\end{eqnarray}
where $\tau$ and $\chi$ are the time variable and
the radial coordinate, respectively. Under a transformation such as
\begin{equation}
u = -\frac{2 \nu}{\theta}
 \frac{\partial \theta}{\partial \chi} \,,
\end{equation}
Eq.~(\ref{eqn:cylindrical-Burgers}) is rewritten as
\begin{equation} \label{eqn:cylindrical-Burgers2}
\partial_{\tau} \theta = \frac{\nu}{\chi}
 \partial_{\chi} \left(\chi \partial_{\chi}\theta \right) \,.
\end{equation}
The generic solution for Eq.~(\ref{eqn:cylindrical-Burgers2})
is described by integral form:
\begin{eqnarray}
\theta(\chi, \tau) &=& k(\tau) \exp \left (-\frac{1}{2\nu}
 \int_0^{\chi} u(\omega, \tau) d \omega \right ) \,, \\
k(\tau) &=& \theta (0, \tau) \,, \\
\theta(0, \tau) &=& \theta_0(\chi)
 = k(0) \exp \left (-\frac{1}{2\nu} 
 \int_0^{\chi} u(\omega, \tau) d \omega \right ) \,.
\end{eqnarray}

\begin{figure}
 \includegraphics[height=7cm]{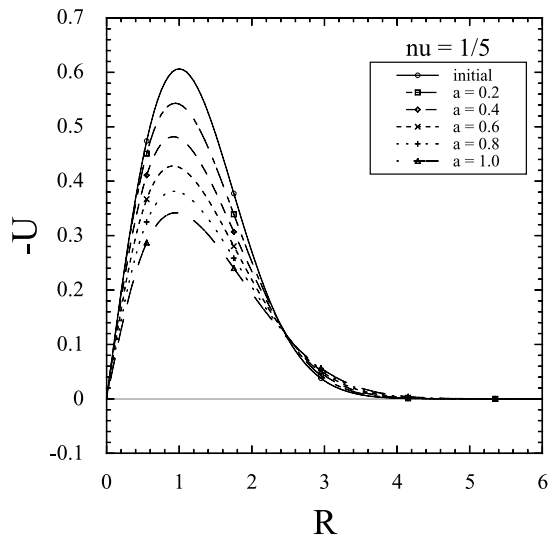}
 \caption{The evolution of the ``peculiar velocity'' in the AA model.
 Here we show the velocity distribution at $a=1/60$ (initial, solid line)
 and $a=0.2, 0.4, \cdots, 1.0$. During evolution, the velocity
 decelerates because the viscosity resists the motion of the fluid.
 In AA, although the velocity decelerates, it does not oscillate.
 }
 \label{fig:AA-evolv}
\end{figure}

In this paper we apply the cylindrical Burgers's
equation~(\ref{eqn:cylindrical-Burgers}) for the adhesion model.
In the E-dS Universe model, the evolution equation is given as
\begin{equation} \label{eqn:Cylindrical-AA}
\frac{\partial u}{\partial a} + u \frac{\partial u}{\partial R}
 = \nu \frac{\partial}{\partial R} \left [ \frac{1}{R}
 \frac{\partial}{\partial R} \left (R u \right ) \right ] \,.
\end{equation}
Now we introduce Lagrangian displacement
for the radial coordinate from Eq.~(\ref{eqn:x=q+s}):
\begin{equation} \label{eqn:x=R+s}
R = r + \partial_r S(r, t) \,,
\end{equation}
where $R$ and $r$ mean Eulerian and Lagrangian radial
coordinate, respectively, and $S$ means Lagrangian displacement potential.
Because we assume cylindrical symmetry and irrotational motion,
the Lagrangian perturbation includes
only the longitudinal mode.
Also the ``peculiar velocity'' (Eq.~(\ref{eqn:def-u})) is rewritten as
\begin{equation}
u = \frac{\partial}{\partial a} \left (
 \frac{\partial S}{\partial r} \right ) \,.
\end{equation}
Using this ``peculiar velocity,'' the evolution equation of the
adhesion model is described by Eq.~(\ref{eqn:Cylindrical-AA}),
and we examine the evolution of the ``peculiar velocity.''
For comparison, we calculate cylindrical models with
the viscous parameter $\nu$.
Figure~\ref{fig:AA-evolv} shows the evolution of
the ``peculiar velocity'' with $\nu=1/5$, where we show the relation between
Eulerian coordinates $R$ and the ``peculiar velocity'' $U$. 
If we take a limit $\nu \rightarrow 0$, i.e, for the case of dust (ZA),
the distribution of the ``peculiar velocity'' at any given time is the
same as the initial distribution. On the other hand,
from Fig.~\ref{fig:AA-evolv} we can see that
the ``peculiar velocity'' decelerates monotonically due to the viscosity.

We notice that Eq.~(\ref{eqn:Cylindrical-AA})
is described with Eulerian comoving coordinates.
Thus it is complicated in a degree if we obtain the Lagrangian
displacement (cf. Eq.~(\ref{eqn:sol-AA})). 
In fact, for evolution of the fluctuation we need to consider
the correspondence between Eulerian coordinates $\bm{x}$ and
Lagrangian coordinates $\bm{q}$ for every
grid at every time-step.

% we need to compute the ``peculiar velocity''
%which corresponds to Lagrangian coordinates $\bm{q}$ at
%every time-step. In other words, although the ``peculiar velocity''
%depends on the Eulerian coordinates $\bm{x}$, the Eulerian
%coordinates is function of the Lagrangian coordinates $\bm{q}$.

%%% Figure %%%
%%%%%%%%%%%%%

%%%%%%%%%%%%%%%%%%%%%%%%%%%%%%%%%%%%%%%%%%%%%%%%%%
\subsection{The EJN model}\label{sec:compari-EJN}
%%%%%%%%%%%%%%%%%%%%%%%%%%%%%%%%%%%%%%%%%%%%%%%%%%

Next, we analyze evolution in the EJN model.
If we consider a cylindrical model, because of mode-coupling in
Laplacian, we cannot separate the perturbation to
the spacial-dependent and the time-dependent term.
Therefore we adopt numerical calculation for the evolution
of the EJN model.
In a previous paper~\cite{Tatekawa04}, 
we apply an explicit method 
for solving partial differential equations
\cite{NumericalRecipes}.
However this method tends to produce
numerical instability. To avoid the numerical
instability, we adopt another method. Teukolsky~\cite{Teukolsky}
discussed the iterated Crank-Nicholson method, which is one
implicit scheme for numerical calculation. The method 
resolves the difficulty of solving the resulting implicit
algebraic equations in the original Crank-Nicholson method
and preserves the good stability properties.

The ``peculiar velocity''
in comoving coordinates is described as
\begin{equation}
\bm{v} (a, \bm{q}) \equiv \frac{\partial}{\partial a} \left (
 \nabla_q S(a, \bm{q}) \right ) \,.
\end{equation}
Because we consider a cylindrical symmetric model, we compute only
radial velocity, i.e.,
\begin{equation}
\bm{V}_r (a, r) \equiv \frac{\partial}{\partial a} \left (
 \frac{\partial S(a, r)}{\partial r} \right )\,,
\end{equation}
where $r$ is the Lagrangian radial coordinate.
From the definition of the ``peculiar velocity,''
we notice it does not change in the dust model (ZA).
Because the solution in the dust model can
be decomposed in time and spacial components, 
the ``peculiar velocity'' at any time is the same
as the initial ``peculiar velocity.''
Furthermore, the growing factor of the dust model
is given by Eq.~(\ref{eqn:ZA-grow}). Therefore the 
longitudinal mode $S$ in ZA monotonously increases
(Eq. (\ref{solution-ZA})).

Correspondingly we analyze the ``peculiar velocity'' in the EJN model.
For the
numerical calculation, we set the boundary condition at $r=0$ and
$r=10$ such as
\begin{eqnarray}
\left . \frac{\partial S(a, r)}{\partial r} \right |_{r=0}
 &=& 0 \,, \\
S(a, 10) = \left . \frac{\partial S(a, r)}{\partial r} \right |_{r=10}
 &=& 0 \,.
\end{eqnarray}
Because the distribution is asymptotically homogeneous, we
set that the fluid does not move at the outside.
The behavior of the EJN model strongly depends on the parameters
$\kappa$ and $\gamma$. If $\kappa$ is very small, the solution
is similar to that in ZA.
On the other hand, if $\kappa$ is
very large, the fluctuation oscillates and disappears.
Following our previous method~\cite{Tatekawa04}, we choose a
reasonable value of the parameter $\gamma$, i.e.,
$\gamma=4/3$ and $5/3$. In the case of $\gamma=4/3$,
as we showed in Eq.~(\ref{hatS43}),
the solutions can be described simply and have both growing and
decaying modes. In the case of $\gamma=5/3$, Buchert and Dom\'{\i}nguez
\cite{budo}
claimed that the isotropic velocity dispersion corresponds
to the origin of the artificial viscosity in AA. Further,
instead of $\kappa$, we set the value of the Jeans wavenumber
$K_J$ (Eq.~(\ref{eqn:Jeans-WN})).

\begin{figure}
 \includegraphics[height=7cm]{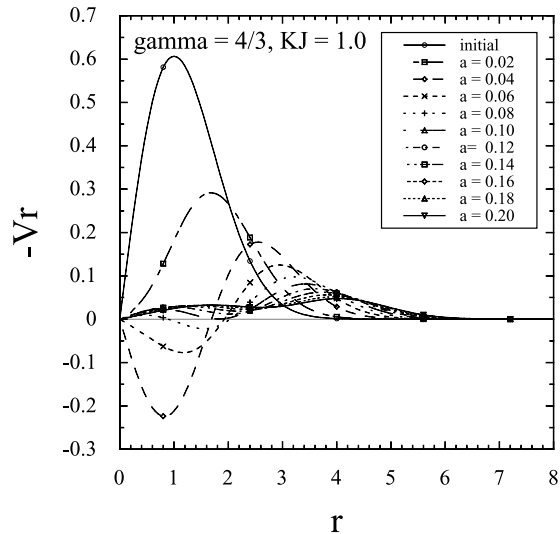}
 \caption{The evolution of the ``peculiar velocity'' in the EJN model
 ($\gamma=4/3, K_J=1.0$).
 We show the time slices at $a=0.02, 0.04, \cdots, 0.2$.
 We also show the velocity distribution at $a=1/60$ (initial, solid line).
 During time evolution, the peak of the ``peculiar velocity'' moves
 to the outside.
 Because of the effect of the pressure,
 the form of the distribution of the ``peculiar velocity'' changes from
 the initial one. The velocity decelerates and the falling fluid at the initial time
 bounces to the outside.
 }
\label{fig:Vr-g43}
\end{figure}

\begin{figure}
 \includegraphics[height=7cm]{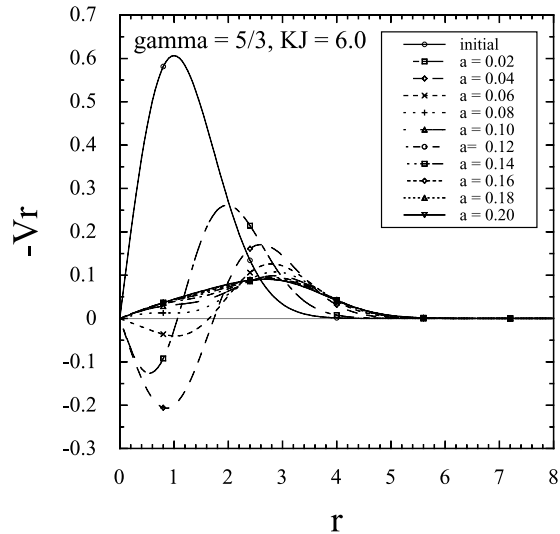}
  \caption{The evolution of the ``peculiar velocity'' in the EJN model
 ($\gamma=5/3, K_J=6.0$).
 We show the time slices at $a=0.02, 0.04, \cdots, 0.2$.
 We also show the velocity distribution at $a=1/60$ (initial, solid line).
 Similar to the case where $\gamma=4/3$,
 during time evolution, the peak of the ``peculiar velocity'' moves
 to the outside. 
 Because of the effect of the pressure,
 the distribution of the ``peculiar velocity'' oscillates at $0.02 < a < 0.2$.
 }
\label{fig:Vr-g53}
\end{figure}

Figures~\ref{fig:Vr-g43} and \ref{fig:Vr-g53} show
the ``peculiar velocity'' in the EJN model with linear approximation.
We can see that the solution has both growing and
decaying modes in the early stage,
and that the position where the ``peculiar velocity''
is the fastest moves outward. 
Then the distribution of the ``peculiar velocity''
changes by the effect of the pressure.
Moreover it is known that in Cartesian coordinates the solution for $\gamma \ge 4/3$
always has a growing mode, and that the ``peculiar velocity'' is not zero
at any time (especially in the case where $\gamma=5/3$
the perturbative solutions asymptotically approach those of
the dust model).
Therefore as seen Figs.~\ref{fig:Vr-g43} and \ref{fig:Vr-g53},
the distribution of the ``peculiar velocity'' slightly oscillates, 
and the peak of the ``peculiar velocity'' moves to the outside
at the early stage.
Then, finally, caustics would form in the EJN model
as in the case of the dust model (ZA).

%
%%% Figure %%%
%%%%%%%%%%%%%
%

%
%%% Figure %%%
%%%%%%%%%%%%%
%

%
%%% Figure %%%
\begin{figure}
 \includegraphics[height=7cm]{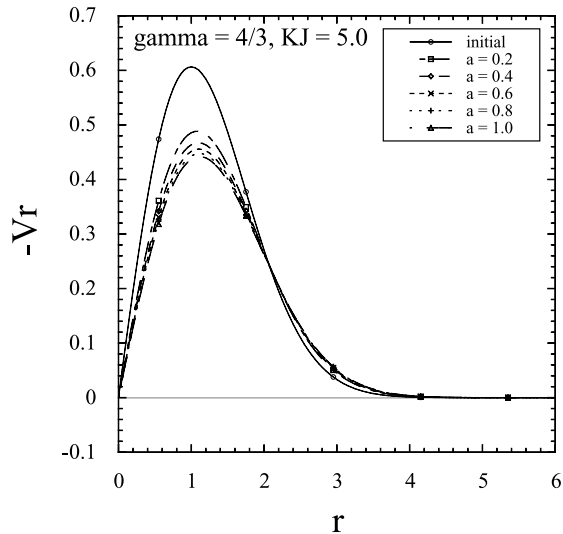}
 \caption{The evolution of the ``peculiar velocity'' in the EJN model
 ($\gamma=4/3, K_J=5.0$).
 We show the time slices at $a=0.2, 0.4, \cdots, 1.0$.
 We also show the velocity distribution at $a=1/60$ (initial, solid line).
 During time evolution, the peak of the ``peculiar velocity''
 slightly moves to the outside.
 In this case, the ``peculiar velocity'' does not oscillate.
 However, the cylindrical dust would collapse, because the deceleration
 of the ``peculiar velocity'' is gentle.
 }
\label{fig:Vr-g43A}
\end{figure}
%%%%%%%%%%%%%
%

%%%%%%%%%%%%%%%%%%%%%%%%%%%%%%%%%%%%%%%%%%%%
\section{Discussion and Concluding Remarks}
\label{sec:discuss}
%%%%%%%%%%%%%%%%%%%%%%%%%%%%%%%%%%%%%%%%%%%%

With respect to the distribution of the ``peculiar velocity,''
we examine the correspondence between AA and the EJN model
with cylindrical symmetry.
In this analysis,
even if we consider linear perturbation, it is hard to
describe the solution with explicit form. Therefore
we carried out numerical calculation, where to avoid
numerical instability, we adopted the iterated
Crank-Nicholson method.
From our calculation, when we take small value for the
Jeans wavenumber,
the ``peculiar velocity'' in the EJN
model oscillates due to the pressure. For the cosmological
fluid, such oscillation is known as the Jeans
instability. On the other hand, the ``peculiar velocity'' in AA
is decelerated due to the viscosity, where there is no oscillation.
Thus we can see that the behavior of AA is different from that of the EJN model.

Furthermore, when we take a large value for the Jeans wavenumber
in the EJN model, the evolution of the ``peculiar velocity''
is similar to that in AA (see Figure~\ref{fig:Vr-g43A}).
However, it is well known that in Cartesian coordinates
the linear perturbative solutions for $\gamma \ge 4/3$ in the EJN model
have a growing solution.
Therefore the perturbation in the EJN model
would eventually diverge. In other words,
we can predict that even if we take a large value for the Jeans
wavenumber in the EJN model, the radial motion of the fluid
does not stop and finally collapses.
Thus there also exists an essential difference between AA and the EJN  model
on this point,
because it is known that the formation of caustic does not occur in AA.
Hence AA cannot express the EJN model with propriety,
and in order to clarify the origin
of the viscosity term in AA, we should consider other
effects besides isotropic velocity dispersion or isotropic
effective pressure.

Recently, Buchert and Dom\'{\i}nguez~\cite{Buchert05} discussed
adhesive gravitational clustering and tried to provide a clear
explanation of the assumuptions for adhesion approximation.
They applied the Eulerian and Lagrangian expansions to 
the non-perturbative regime and proposed a new non-perturbative
approximation. When we analyze this new approach with both
analytic and numerical methods, we may be able to explicate the origin
of the artificial viscosity in AA. Also, as a future work,
we would describe the evolution of the density fluctuation in a
nonlinear regime with a semi-analytic method.

\begin{acknowledgments}
We thank Masahiro Morikawa for useful discussion.
TT thanks the Mathematics Library, Hokkaido University,
for pointing out some references.
This work was supported by the Grant-in-Aid for Scientific
Research Fund of the Ministry of Education, Culture, Sports, Science
and Technology, Japan (Young Scientists (B) 16740152).
\end{acknowledgments}

\appendix

%%%%%%%%%%%%%%%%%%%%%%%%%%%%%%%%%%%%%%%%%%%%%%%%%%%%%%%%
\section{The evolution equation in cylindrical model}
\label{sec:Sphe-eqs}
%%%%%%%%%%%%%%%%%%%%%%%%%%%%%%%%%%%%%%%%%%%%%%%%%%%%%%%%

In Sec.~\ref{subsec:pressure}, we noticed that Eq.~(\ref{eqn:P-full})
is hard to solve because of the coordinate transformation or non-locality.
In this appendix we rewrite the equation with cylindrical symmetry.
We can introduce Lagrangian displacement
for the radial coordinate by Eq.~(\ref{eqn:x=R+s}).
The spacial derivative is rewritten with the Lagrangian coordinate, such as
\begin{eqnarray}
\frac{\partial}{\partial r} = \frac{\partial R}{\partial r}
 \frac{\partial}{\partial R}
 = \left ( 1 + \partial_{r}^2 S \right ) \frac{\partial}{\partial R} \,.
\end{eqnarray}
Therefore, the derivative is changed as
\begin{equation}
\frac{\partial}{\partial R} = \frac{1}{1+ \partial_{r}^2 S}
\frac{\partial}{\partial r} \,.
\end{equation}
The divergence of the ``peculiar velocity'' with a Eulerian coordinate
becomes a little complicated:
\begin{eqnarray}
\nabla_x \cdot {\bf u}
 &=& \frac{1}{R} \frac{\partial}{\partial R} \left (R \partial_{R} u \right )
 \nonumber \\
 &=& \frac{1}{(r+ \partial_r S)} \frac{1}{1+ \partial_{r}^2 S}
 \frac{\partial}{\partial r} \left \{
 (r+ \partial_{r} S) \partial_r u \right \} \,.
\end{eqnarray}

We decompose the Jacobian of the coordinate transformation 
to the order of the perturbation.
\begin{equation}
J=1+J^{(1)}+J^{(2)}+J^{(3)} \,,
\end{equation}
where
\begin{eqnarray}
J^{(1)} &=& \partial_i^2 S = \partial_r^2 S + \frac{1}{r} \partial_r S \,, \\
J^{(2)} &=& \frac{1}{2} \left [
 \left(\partial_i^2 S \right) \left(\partial_j^2 S \right)
 - \left(\partial_i \partial_j S \right) \left(\partial_j \partial_i S \right) \right ]
\nonumber \\
 &=& \frac{1}{r} \left(\partial_r S \right) \left(\partial_r^2 S \right) \,, \\
J^{(3)} &=& \det \left (\partial_{i}\partial_j S \right )
 = 0 \,.
\end{eqnarray}
Most of above changes affect only higher-order approximation.
In the EJN model, we must consider the change of Laplacian
in Lagrangian space, i.e.,
\begin{eqnarray}
\nabla_q^2 f(r) &=& \frac{1}{r} \frac{\partial}{\partial r}
 \left (r \frac{\partial f(r)}{\partial r} \right ) \nonumber \\
&=& \frac{1}{r} \left ( f'(r) + r f''(r) \right ) \,,
\end{eqnarray}
where the prime denotes the partial derivative with respect to $r$.
Using the above deformations, we obtain the evolution equation
for the EJN model with cylindrical symmetry.
The linear perturbative equation for the EJN model is
written as
\begin{equation} \label{eqn:EJN-cylindrical1}
\frac{1}{r} \frac{\partial}{\partial r} \left [ r
 \left \{ \ddot{S}'
 + 2 \frac{\dot{a}}{a} \dot{S}'
 - \frac{\kappa \gamma \rho_b^{\gamma-1}}{a^2}
  \frac{\partial}{\partial r} \left (
 \frac{1}{r} \frac{\partial}{\partial r} \left ( r
 S' \right ) \right )
 \right \} \right ] = -4 \pi G \rho_b \cdot
 \frac{1}{r} \frac{\partial}{\partial r} \left [ r
 S' \right ] \,.
\end{equation}
When we choose an appropriate boundary condition, we can
rewrite Eq.~(\ref{eqn:EJN-cylindrical1}) as
\begin{equation} \label{eqn:EJN-cylindrical2}
\ddot{S} +  2 \frac{\dot{a}}{a} \dot{S}
 -  \frac{\kappa \gamma \rho_b^{\gamma-1}}{a^2} 
 \left (S'' + \frac{1}{r} S' \right )
 = -4 \pi G \rho_b S \,.
\end{equation}
When we treat the EJN model in Cartesian coordinates, we can
apply a Fourier transformation to the perturbative equations easily.
However, with the cylindrical symmetry, because
the fourth term on the left-hand side
in Eq.~(\ref{eqn:EJN-cylindrical2}) derives convolution,
even if we take only linear perturbation, we cannot obtain
analytic solutions and we have to solve by numerical calculation.
\end{document}